\documentclass[
reprint,
aps,pra,
superscriptaddress,
amsmath,amssymb,
floatfix,
longbibliography,
nofootinbib
]{revtex4-2}

\allowdisplaybreaks

\usepackage{graphicx}
\usepackage{dcolumn}
\usepackage{bm}
\usepackage{etoolbox}

\usepackage[utf8]{inputenc}
\usepackage[T1]{fontenc}
\usepackage[english]{babel}

\usepackage{amsthm}

\usepackage[caption=false]{subfig}
\usepackage{tikz}

\providecommand{\orcid}[1]{}

\providecommand{\dd}{\mathop{}\!\mathrm{d}}

\usepackage{xurl}
\usepackage{hyperref}
\AtBeginEnvironment{thebibliography}{\sloppy}

\usepackage{placeins}

\begin{document}

\title{Collective Mpemba-Type Relaxation in Degenerate Bosonic Modes Coupled
	to a Common Thermal Reservoir}

\author{I. M. de Araújo}
\affiliation{Instituto de Física de São Carlos, Universidade de São Paulo, Caixa Postal 369, 13560-970, São Carlos, São Paulo, Brazil}
\thanks{Corresponding author: \href{mailto:italo.maraujo@ifsc.usp.br}{italo.maraujo@ifsc.usp.br}}
\orcid{0000-0002-8760-7731}

\author{H. Sanchez}
\affiliation{Instituto de Física de São Carlos, Universidade de São Paulo, Caixa Postal 369, 13560-970, São Carlos, São Paulo, Brazil}
\orcid{0000-0002-4915-4495}

\author{L. F. Alves da Silva}
\affiliation{Instituto de Física de São Carlos, Universidade de São Paulo, Caixa Postal 369, 13560-970, São Carlos, São Paulo, Brazil}
\orcid{0000-0002-8418-4833}

\author{Norton G. de Almeida}
\affiliation{Instituto de Física, Universidade Federal de Goiás, Goiânia, Goiás, Brazil}

\author{M. H. Y. Moussa}
\affiliation{Instituto de Física de São Carlos, Universidade de São Paulo, Caixa Postal 369, 13560-970, São Carlos, São Paulo, Brazil}
\orcid{0000-0002-3026-0845}

\begin{abstract}
We investigate collective Mpemba-type relaxation in a degenerate family of bosonic modes coupled to a common thermal reservoir. Starting from a fully symmetric $M$-mode description and employing a representative mean-field reduction, we derive effective master equations for weak-coupling Markovian, strong-coupling Caldeira–Leggett, and weak-coupling non-Markovian regimes. In the weak-coupling Markovian limit, relaxation separates into an incoherent thermal channel decaying at rate $\gamma$ and a collective coherent channel decaying at rate $M\gamma$, yielding an explicit Mpemba crossing time determined by the initial-state preparation. In the strong-coupling Caldeira–Leggett regime, transient quadrature dynamics enriches the relaxation pattern, delaying crossings in the overdamped sector and generating multiple crossings in the underdamped sector. In the non-Markovian regime, reservoir memory reshapes the same channel-competition mechanism through time-dependent decay rates and a Lamb shift, producing delayed and clustered crossing windows. Numerical results based on the energy and the Kullback–Leibler divergence reveal Mpemba and anti-Mpemba behavior, criterion dependence, and multiple transient reorderings induced by collective coherence, quadrature dynamics, and reservoir memory.
\end{abstract}
\maketitle

\section{Introduction}


The Mpemba effect refers to the counterintuitive phenomenon whereby
a system initially at a higher temperature than the thermal reservoir
reaches equilibrium sooner than an otherwise identical system initially at a
lower temperature. The idea goes back to freezing experiments with water 
\cite{MpembaOsborne1969}, but later work made clear that, in macroscopic
systems, its observation depends strongly on protocol, since convection,
evaporation, supercooling, and boundary conditions all compete \cite%
{BurridgeLinden2016}.

Mechanism-oriented modeling has shown, for example, how subtle differences
in preparation can bias nucleation, local structure, and heat transport \cite%
{JinGoddard2015}, and how evaporation and natural convection reshape cooling
curves under realistic boundary conditions \cite%
{VynnyckyMitchell2010,VynnyckyKimura2015}. Taken together, these
lessons make clear that Mpemba-type statements are ultimately statements
about relaxation times, under a specified protocol and relative to a
specified criterion.

A natural way to separate mechanism from protocol is to fix the dynamical
generator and ask how different initial preparations distribute their weight
among the distinct contributions that govern relaxation. In Markovian
regimes, this viewpoint connects directly to the interpretation of Mpemba
and inverse-Mpemba effects in terms of different weights on the slow and
fast components of relaxation \cite{LuRaz2017,HoltzmanRaz2022}. It also
reflects the fact that many experimental claims are most naturally framed in
terms of the time needed for a given effect to become observable.

This perspective becomes even more compelling in systems with memory or with
multiple nonequilibrium resources. Beyond simple liquids, memory effects in
complex energy landscapes provide an instructive classical parallel \cite%
{BaityJesi2019}. In quantum systems, the problem is sharpened by the fact
that nonequilibrium can be stored in different resources, most notably
populations and coherences \cite{AresCalabreseMurciano2025}. Mpemba-type
phenomena have been analyzed in closed dynamics \cite{Rylands2024} and
observed in controlled quantum simulations \cite{Joshi2024}; in open quantum
systems they have been explored in nonequilibrium regimes \cite%
{NavaEgger2024,WangWang2024}. Photonic and laser platforms provide
additional controlled realizations \cite%
{Longhi2024Photonic,Longhi2025Laser,LonghiQuantum2025}.

Recent syntheses point to a common theme across these developments: many
manifestations can be traced to competing contributions to a chosen measure
of proximity to equilibrium that decay on different time scales \cite%
{TezaBechhoeferLasantaRazVucelja2026}. Related viewpoints likewise emphasize
that both the relaxation spectrum and the diagnostic used to define
proximity to equilibrium can play a decisive role \cite%
{YuLiuZhang2025,LiuZhang2024RandomCircuits,ChatterjeeTakadaHayakawa2024}.

In this work, we investigate how the competition between incoherent and coherent relaxation channels is modified by collective dissipation in a degenerate set of bosonic modes coupled to a common thermal reservoir. We analyze three representative situations: weak coupling to a Markovian bath, strong coupling to a Markovian bath described at the Caldeira–Leggett level, and weak coupling to a genuinely non-Markovian reservoir. Throughout, we start from a collective M-mode description and derive representative master equations by tracing out $M-1$ modes under permutation symmetry and a mean-field closure.

The collective character of the model originates from the fact that all degenerate modes couple to the same reservoir channel. As a consequence, the dissipative dynamics contains crossed reservoir-induced contributions that reorganize the degenerate manifold into collective bright and dark sectors, in close analogy with superradiant and subradiant relaxation mechanisms. The common bath therefore generates relaxation scales that depend explicitly on the number of modes M, so that the coherent component associated with the bright sector may relax collectively faster than the incoherent thermal contribution.

Within this framework, the weak-coupling Markovian regime provides an analytically solvable reference model in which the time-independent generator makes the origin of Mpemba crossings especially transparent. In this case, the collective bright sector generates a coherent contribution relaxing at an enhanced rate proportional to $M\gamma$, yielding explicit collective modifications of the Mpemba crossing timescale. We then extend the same representative strategy in two directions: first, to a Caldeira–Leggett description in the strong-damping regime; second, to a genuinely non-Markovian reservoir with time-dependent coefficients and a Lamb shift. Taken together, these regimes highlight two complementary physical ingredients. In both the weak-coupling Markovian and non-Markovian regimes, the relevant mechanism is the competition between incoherent and coherent contributions, while in the non-Markovian case reservoir memory reshapes the effective relaxation timescale through non-exponential damping and time-dependent relaxation rates. In the Caldeira–Leggett regime, by contrast, the key new ingredient is transient quadrature dynamics, a feature absent from both weak-coupling cases, which can either delay a crossing or split it into several crossings over a finite time window.

We stress that the collective nature of the model does more than simply renormalize the relaxation scales. Because all modes interact through the same environmental channel, the reservoir selectively damps the collective bright sector while leaving orthogonal dark combinations comparatively less affected. As a result, the parameter $M$ controls not only the speed of the coherent relaxation channel, but also the qualitative structure of the nonequilibrium dynamics itself. Depending on the dynamical regime, collective dissipation can accelerate coherent relaxation, delay crossings through overdamped quadrature dynamics, or generate repeated reordering events through underdamped oscillations and memory-induced relaxation reshaping.

\section{Degenerate bosonic modes weakly coupled to a Markovian reservoir}

\label{sec:gksl} 

We begin with a Hamiltonian given by $H=H_{0}+H_{I}$, where (in units $\hbar
=1$):  
\begin{eqnarray}
	H_{0} &=&\omega _{0}\sum_{\ell =1}^{M}a_{\ell }^{\dagger }a_{\ell
	}+\sum_{k}\omega _{k}b_{k}^{\dagger }b_{k},\qquad  \\
	H_{I} &=&\sum_{\ell ,k}\chi _{\ell k}\left( a_{\ell }b_{k}^{\dagger
	}+a_{\ell }^{\dagger }b_{k}\right) ,
\end{eqnarray}%
where $H_{0}$ describes the set of degenerate bosonic modes, of frequency $%
\omega _{0}$, and a multimodal bosonic reservoir, of frequencies $\left\{
\omega _{k}\right\} $. $H_{I}$ describes the weak coupling between the
degenerate modes and the reservoir degrees of freedom. The master equation
describing the dynamics of the system--reservoir coupling is given by
\begin{equation}
\begin{aligned}
	\dot{\rho}_{M}
	={}& -i\left[\omega _{0}\sum_{\ell =1}^{M}a_{\ell }^{\dagger }a_{\ell},
	\rho _{M}\right] \\
	&+\gamma (n_{B}+1)\,\mathcal{L}[A]\rho _{M}
	+\gamma n_{B}\,\mathcal{L}[A^{\dagger }]\rho _{M}.
\end{aligned}
\end{equation}%
where $\gamma$ is the strength of the system--reservoir coupling, $%
n_{B}=\left( e^{\beta \omega _{0}}-1\right) ^{-1}$ is the reservoir
excitation at frequency $\omega _{0}$, with $\beta =1/k_{B}T$, $k_{B}$ being
the Boltzmann constant and $T$ the temperature of the reservoir. Moreover,
by defining the collective operator $A=\sum_{\ell =1}^{M}a_{\ell }$, the
emission and absorption Lindbladians read  
\begin{subequations}
	\label{eq:Lindblad_collective}
	\begin{align}
		\mathcal{L}[A]\rho_M
		&= 2A\rho_M A^{\dagger}
		-\{A^{\dagger}A,\rho_M\},
		\label{eq:Lindblad_collective_a}
		\\
		\mathcal{L}[A^{\dagger}]\rho_M
		&= 2A^{\dagger}\rho_M A
		-\{AA^{\dagger},\rho_M\}.
		\label{eq:Lindblad_collective_b}
	\end{align}
\end{subequations}

The collective character of the weak-coupling Markovian dynamics is already encoded in the structure of the jump operator $A=\sum_{\ell=1}^{M}a_{\ell}$. Expanding the dissipator therefore generates not only local decay terms, but also cross-damping contributions coupling different modes. These crossed Lindblad terms are a direct consequence of the fact that all degenerate modes interact with the same reservoir channel, so that the environment does not distinguish which mode emitted or absorbed an excitation. Equivalently, introducing the normalized bright mode $B=M^{-1/2}\sum_{\ell=1}^{M}a_{\ell}=A/M^{1/2}$, one finds that the collective dissipator acts only on this bright combination, with an enhanced relaxation scale proportional to $M\gamma$, while the remaining orthogonal combinations behave as dark modes with respect to the common bath. In this sense, the relevant collective ingredient is not merely the presence of $M$ independent oscillators, but the reservoir-induced reorganization of the degenerate manifold into bright and dark sectors. The $M$-dependent relaxation scale obtained below should therefore be understood as the relaxation of the bright coherent component selected by the common reservoir.

In order to address the Mpemba effect resulting from the weak coupling of
degenerate modes with the Markovian reservoir, we proceed with the
mean-field approximation, by which we reduce the set of modes to a
representative one. In computing the reduced density operator for the
representative mode ---i.e., by tracing out the remaining $M-1$ modes--- we
thus obtain 
\begin{equation}
	\dot{\rho}=-i\,[H_{\mathrm{eff}}(t),\rho ]+\gamma (n_{B}+1)\,\mathcal{L}[a]\rho
	+\gamma n_{B}\,\mathcal{L}[a^{\dagger }]\rho ,  \label{eq:ME_GKSL}
\end{equation}%
where the instantaneous effective nonlinear mean-field Hamiltonian is given by 
\begin{equation}
	H_{\mathrm{eff}}(t)=\omega _{0}a^{\dagger }a+\frac{i\gamma }{2}(M-1)\left(
	\left\langle a^{\dagger }(t)\right\rangle \,a-\left\langle a(t)\right\rangle
	\,a^{\dagger }\right) .  \label{eq:Heff_GKSL}
\end{equation}

Considering the representative mode prepared in the displaced thermal states 
$\rho (0)=D(\alpha )\rho _{\mathrm{th}}(T_{0})D^{\dagger }(\alpha )$, where $%
D(\alpha )=e^{\alpha a^{\dagger }-\alpha ^{\ast }a}$ is the displacement
operator and $\rho _{\mathrm{th}}(T_{0})=e^{-\beta \omega _{0}a^{\dagger }a}/%
\operatorname{Tr}e^{-\beta \omega _{0}a^{\dagger }a}$ the thermal state at
temperature $T_{0}$ , we obtain its initial excitation $\left\langle
n(0)\right\rangle =n_{0}+|\alpha |^{2}$ where $n_{0}=\left( e^{\beta
	_{0}\omega _{0}}-1\right) ^{-1}$. Moreover, it follows that $\left\langle
a(t)\right\rangle =\,\alpha e^{-(i\omega _{0}+M\gamma /2)t}$ and%
\begin{equation}
	\left\langle n(t)\right\rangle =n_{B}+(n_{0}-n_{B})e^{-\gamma t}+|\alpha
	|^{2}e^{-M\gamma t}.  \label{eqref}
\end{equation}
This last expression makes two contributions explicit: an incoherent
population decay at rate $\gamma $ and a coherent collective decay at rate $%
M\gamma $. For later comparison, we denote the total energy of the $M$
degenerate modes simply by $E=M\omega _{0}\,\left\langle n(t)\right\rangle$. For two different state preparations $\rho _{j}(0)=D(\alpha _{j})\rho _{\mathrm{th}%
}(T_{j})D^{\dagger }(\alpha _{j})$, with $j=1,2$, the energy
difference becomes%
\begin{equation}
	\Delta E(t)\equiv E_{1}(t)-E_{2}(t)=M\omega _{0}\left( \Delta n_{0}\,%
	\mathrm{e}^{-\gamma t}+\Delta \alpha \,\mathrm{e}^{-M\gamma t}\right)
	,\qquad   \label{eq:DeltaE_markov}
\end{equation}%
where $\Delta n_{0}=n_{0,1}-n_{0,2}$ and $\Delta \alpha =|\alpha _{1}|^{2}-|\alpha _{2}|^{2}$. The crossing of the energy decay curves that
characterizes the Mpemba effect ---where the curve corresponding to the
higher initial energy reaches thermal equilibrium before that corresponding
to the lower initial energy--- requires $\Delta n_{0}$ and $\Delta \alpha $
to have opposite signs, together with $-\Delta \alpha /\Delta n_{0}>1$. When
this condition is satisfied, the crossing time is given by 
\begin{equation}
	t_{\times} =\frac{1}{(M-1)\gamma }\ln \!\left( -\frac{\Delta \alpha }{\Delta n_{0}}%
	\right) ,  \label{eq:tx_markov}
\end{equation}%
which provides a concrete realization of the channel-competition picture 
\cite{LuRaz2017,TezaBechhoeferLasantaRazVucelja2026}.

Alongside the energy, we also characterize the Mpemba effect by tracking a
distributional diagnostic, namely the Kullback--Leibler (KL) divergence
between the instantaneous number distribution $p_{t}(n)$ and the thermal
reference $p_{\mathrm{th}}(n)$, as defined by 
\begin{equation}
	D_{\mathrm{KL}}(p_{t}\Vert p_{\mathrm{th}})=\sum_{n\geq 0}p_{t}(n)\,\ln 
	\frac{p_{t}(n)}{p_{\mathrm{th}}(n)}\geq 0,
\end{equation}%
which vanishes if and only if $p_{t}=p_{\mathrm{th}}$, although it is not a
metric \cite{AresCalabreseMurciano2025,TezaBechhoeferLasantaRazVucelja2026}.
In contrast to the energy criterion, which is based solely on the decay of
the mean energy, the Kullback--Leibler-divergence criterion probes the full
nonequilibrium distance to the thermal state, thus providing a more
stringent characterization of the Mpemba effect. This difference simply
reflects the fact that Mpemba-type statements are criterion dependent \cite%
{TezaBechhoeferLasantaRazVucelja2026}.

\subsection{Representative weak-coupling Markovian scenarios}

\begin{figure}[!htb]
	\centering
	\subfloat[Energy $E$.]{%
		\includegraphics[width=0.48\linewidth]{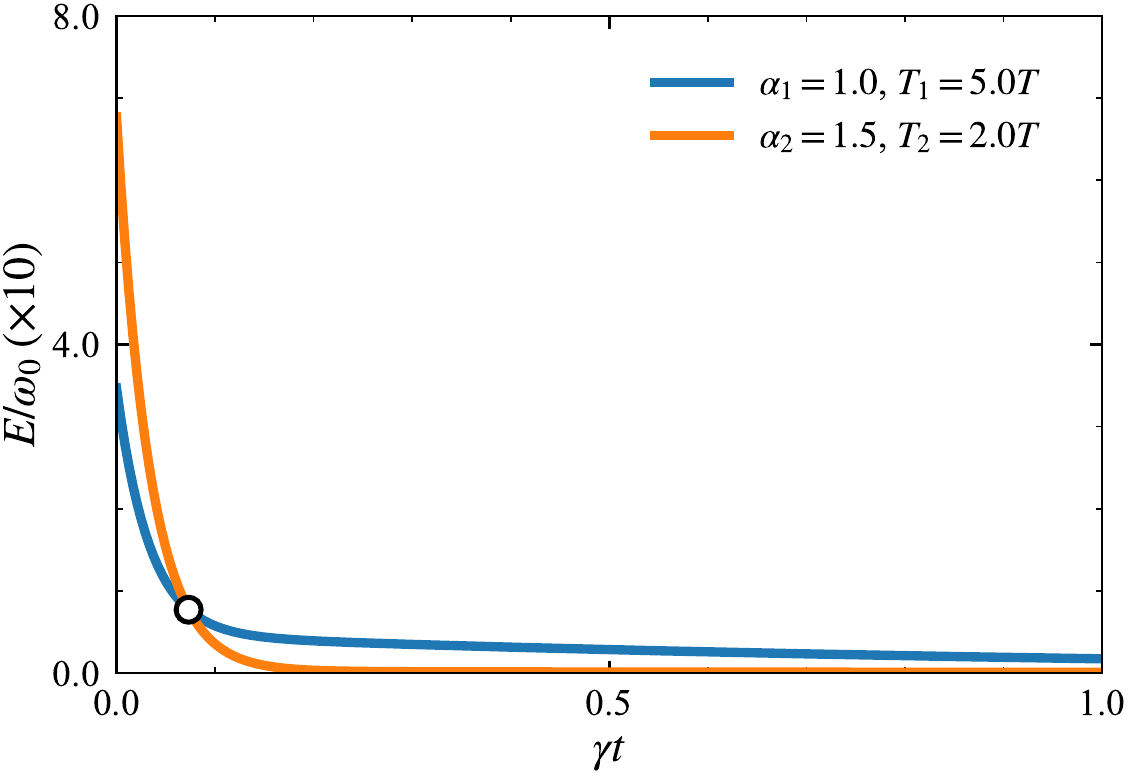}%
	}\hfill\subfloat[KL divergence.]{%
		\includegraphics[width=0.48\linewidth]{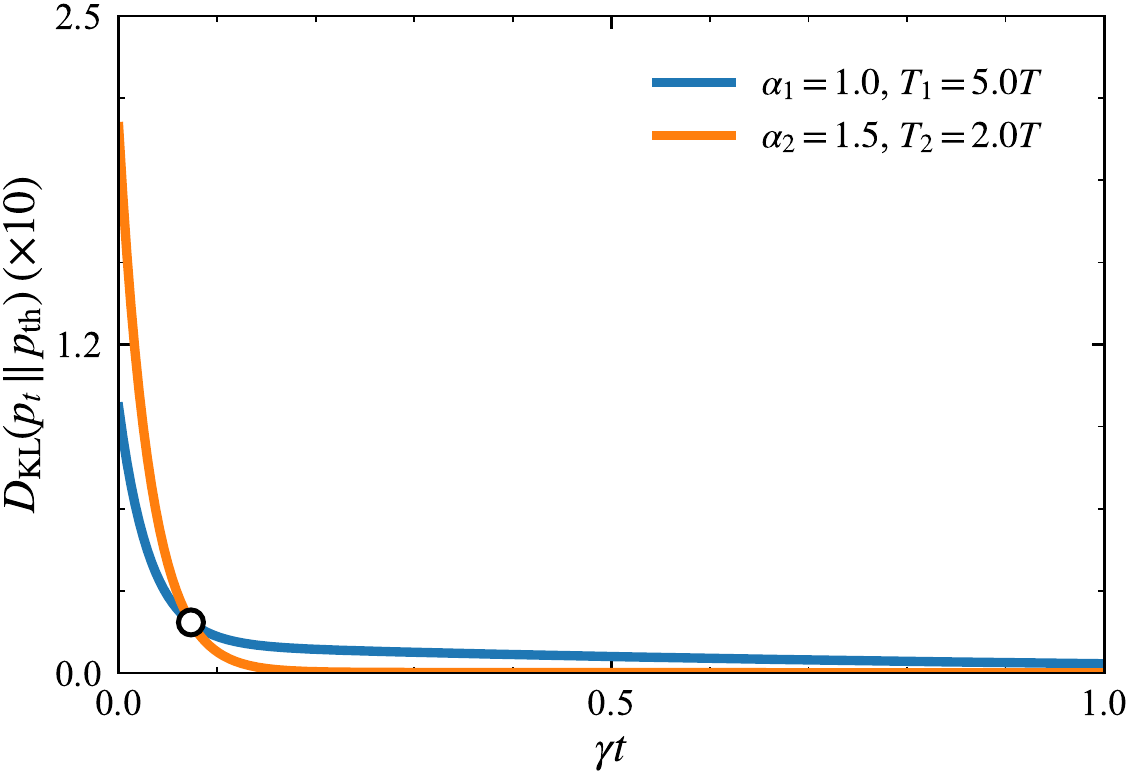}%
	}
	\caption{Weak-coupling Markovian Mpemba scenario with a single clean
		crossing. Parameters: $\omega_0=10$, $\gamma=1$,
		$(\alpha_1,T_1)=(1.0,5.0T)$, and $(\alpha_2,T_2)=(1.5,2.0T)$. The stronger
		initial displacement of state $2$ accelerates the short-time collective
		relaxation and produces the same crossing in both diagnostics.}
	\label{fig:gksl_C1}
\end{figure}

\begin{figure}[!htb]
	\centering
	\subfloat[Energy $E$.]{%
		\includegraphics[width=0.48\linewidth]{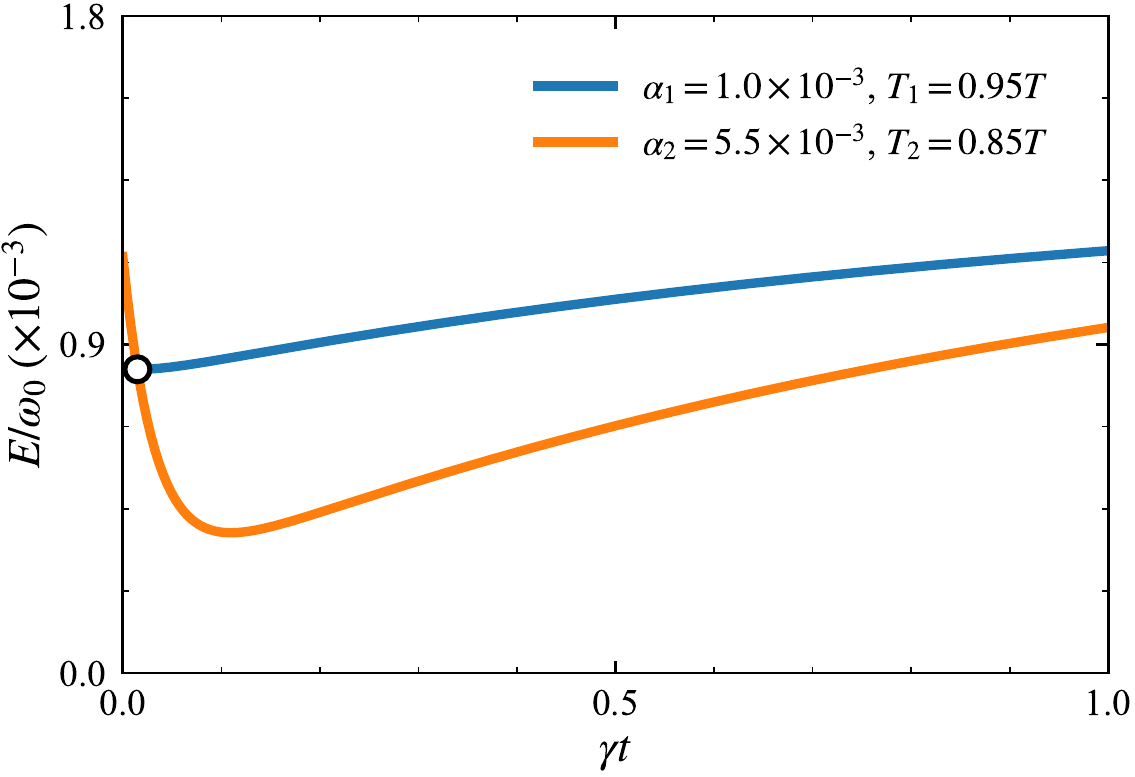}%
	}\hfill\subfloat[KL divergence.]{%
		\includegraphics[width=0.48\linewidth]{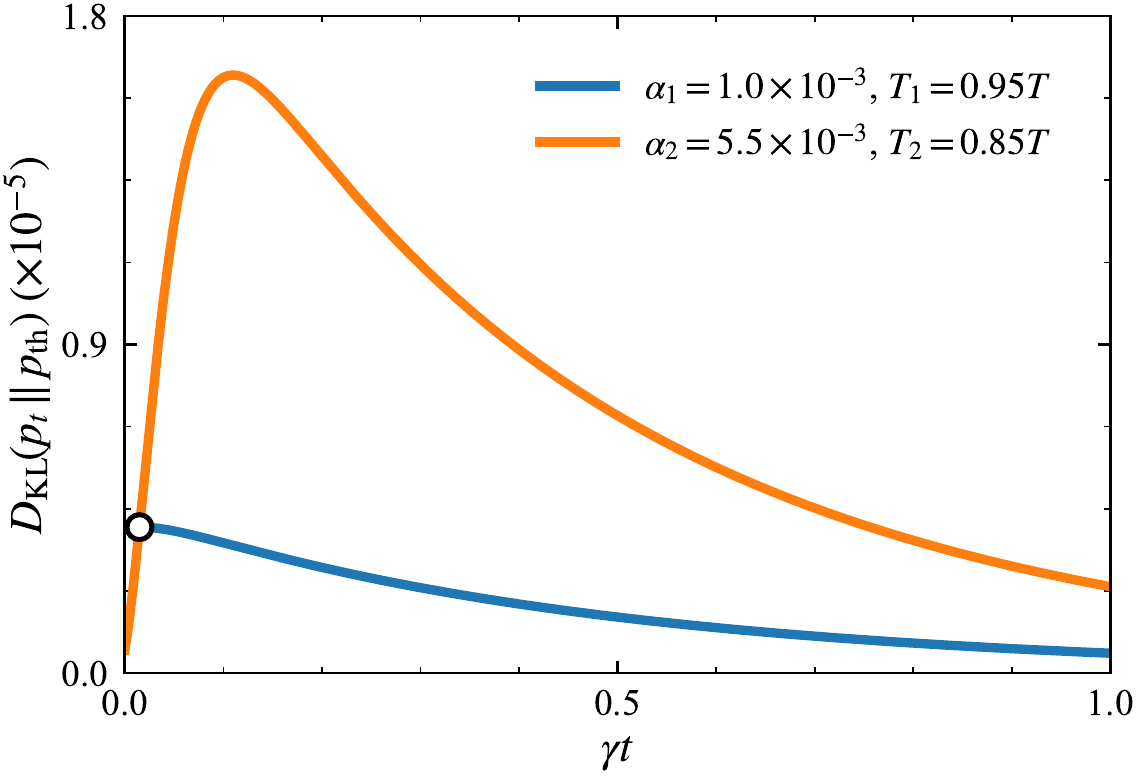}%
	}
	\caption{Weak-coupling Markovian anti-Mpemba-like inversion for
		$\omega_0=10$, $\gamma=1$, $(\alpha_1,T_1)=(10^{-3},0.95T)$, and
		$(\alpha_2,T_2)=(5.5\times 10^{-3},0.85T)$. Although both preparations are
		initially colder than the bath, the larger coherent component of state $2$
		produces an early-time reordering in both the energy and the KL
		divergence.}
	\label{fig:gksl_C2}
\end{figure}

\begin{figure}[!htb]
	\centering
	\subfloat[Energy $E$.]{%
		\includegraphics[width=0.48\linewidth]{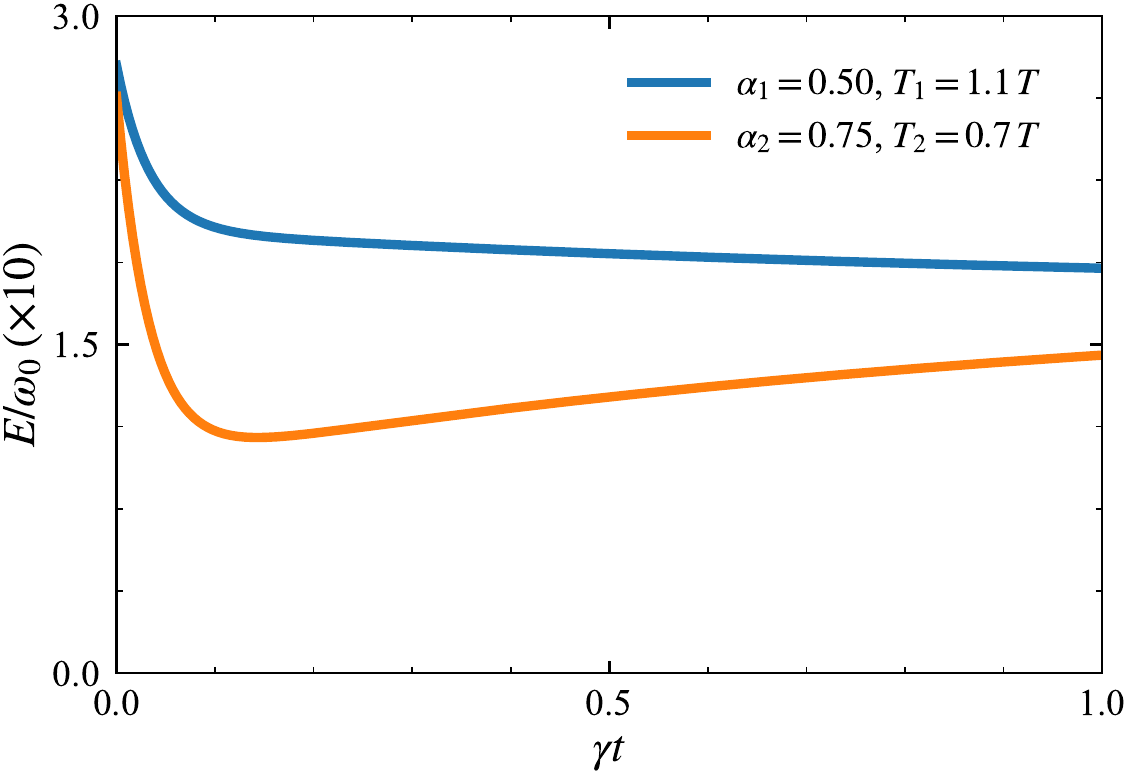}%
	}\hfill\subfloat[KL divergence.]{%
		\includegraphics[width=0.48\linewidth]{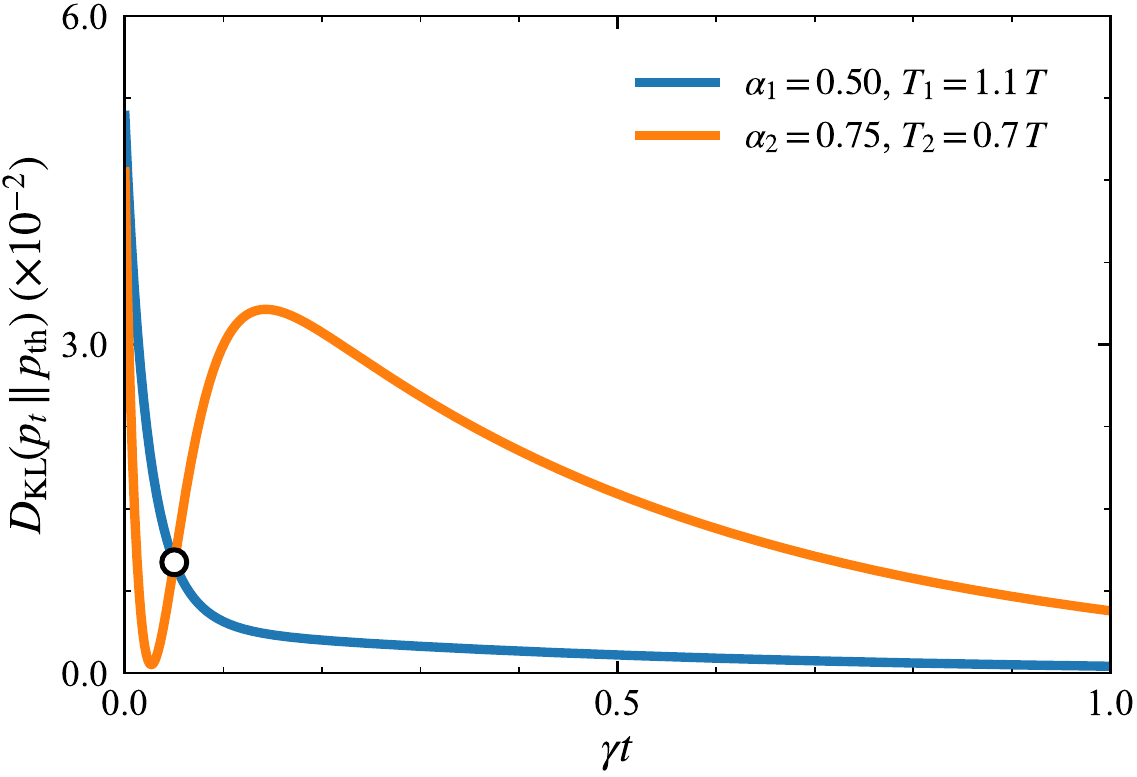}%
	}
	\caption{Weak-coupling Markovian criterion dependence. Parameters:
		$\omega_0=1$, $\gamma=0.1$, $(\alpha_1,T_1)=(0.50,1.1T)$, and
		$(\alpha_2,T_2)=(0.75,0.7T)$. The energy ordering never inverts, whereas
		the KL divergence shows an early crossing, demonstrating that the
		identification of the ``faster'' preparation depends on the chosen
		diagnostic.}
	\label{fig:gksl_C3}
\end{figure}

Figures~\ref{fig:gksl_C1}--\ref{fig:gksl_C3} collect three representative
weak-coupling Markovian examples. In all three cases, the temperatures shown
in the legends are measured in units of the bath temperature $T$, so that,
for instance, $T_1=5.0T$ means a preparation five times hotter than the
reservoir. This is the most transparent regime for interpreting the
collective Mpemba mechanism because Eq.~\eqref{eq:DeltaE_markov} separates
the slow incoherent contribution, proportional to $\mathrm{e}^{-\gamma t}$,
from the faster collective coherent contribution, proportional to
$\mathrm{e}^{-M\gamma t}$. The relative sign and magnitude of $\Delta n_0$
and $\Delta \alpha$ then decide whether the two trajectories preserve their
initial ordering, cross once, or invert only according to the full-state
diagnostic.

Figure~\ref{fig:gksl_C1} shows a clean collective Mpemba crossing for
$\omega_0=10$, $\gamma=1$, $(\alpha_1,T_1)=(1.0,5.0T)$, and
$(\alpha_2,T_2)=(1.5,2.0T)$. Both preparations start hotter than the bath, but
the second one carries a stronger coherent contribution. That larger initial
displacement makes the orange curve drop more rapidly at short times, so the
two trajectories cross once, at about $\gamma t\approx 0.07$, in both the
energy and the KL divergence. After the crossing, the state prepared with
$(\alpha_2,T_2)=(1.5,2.0T)$ remains closer to equilibrium according to both
criteria.

Figure~\ref{fig:gksl_C2} illustrates an anti-Mpemba-like inversion for
$\omega_0=10$, $\gamma=1$, $(\alpha_1,T_1)=(10^{-3},0.95T)$, and
$(\alpha_2,T_2)=(5.5\times 10^{-3},0.85T)$. Here both states are initially
colder than the bath, so the interesting point is not a literal
``hotter-cools-faster'' comparison. Instead, the state with the larger
coherent seed, $\alpha_2=5.5\times 10^{-3}$, undergoes a sharper short-time
reorganization. The energy traces cross almost immediately, and the KL panel
shows the same early reordering even more clearly: the orange trajectory is
briefly the closer one and then becomes the farther one before relaxing back
again.

Figure~\ref{fig:gksl_C3} makes the criterion dependence explicit. In this
case $\omega_0=1$, $\gamma=0.1$, $(\alpha_1,T_1)=(0.50,1.1T)$, and
$(\alpha_2,T_2)=(0.75,0.7T)$. The energy traces never cross: the blue state
remains energetically farther from equilibrium throughout the evolution.
The KL divergence, however, does show an early inversion. Physically, the
larger displacement in the orange state reshapes the full occupation-number
distribution more strongly than it changes the first moment, so the
full-state distance to equilibrium can invert even while the mean-energy
ordering stays fixed.

\FloatBarrier

\section{Degenerate bosonic modes strongly coupled to a Markovian reservoir}

\label{subsec:cl}

We now turn to the case where the degenerate modes are strongly coupled to a
Markovian reservoir, as described by the Caldeira--Leggett linear response
model \cite{CL}. Our starting point is the following equation for the $M$%
-mode state $\rho _{M}$, 
\begin{widetext}
\begin{equation}
	\dot{\rho}_{M}=-i\!\left[\omega_{0}\sum_{\ell=1}^{M}a_{\ell}^{\dagger}a_{\ell},\rho_{M}\right]
	-\frac{i\gamma}{2}\,[X,\{P,\rho_{M}\}]
	-\frac{\gamma}{2}\coth\!\left(\frac{\omega_{0}}{2k_{B}T}\right)[X,[X,\rho_{M}]].
\end{equation}
\end{widetext}%
Here $X=A+A^{\dagger }$, $P=(A-A^{\dagger })/i$, while $A=\sum_{i=1}^{M}a_{i}
$ and $T$ denotes the reservoir temperature. This equation is the collective counterpart of the standard Caldeira--Leggett description of
a harmonic system linearly coupled to a bath of oscillators \cite{CL,CL2,FV}%
. At a schematic level, one begins from the system--reservoir Hamiltonian,
integrates out the environmental degrees of freedom through the
Feynman--Vernon influence-functional approach, and then adopts the
Ohmic and Markovian reduction leading to the Caldeira--Leggett
structure used here, without invoking a further high-temperature
expansion in the coefficients. The resulting equation
contains two characteristic terms: dissipative and diffusive
contributions described by $[X,\{P,\rho _{M}\}]$ and $[X,[X,\rho _{M}]]$,
respectively, the latter encoding reservoir-induced noise.

Applying the same representative-master equation strategy as before
---tracing out $M\!-\!1$ modes and closing the reduced dynamics at the
mean-field level--- one obtains the representative equation 
\begin{equation}
	\begin{aligned}
	\dot{\rho}
	&= -i\,[H_{\mathrm{eff}}(t),\rho]
	-
	\frac{i\gamma}{2}\,[x,\{p,\rho\}] \\
	&\quad
	-
	\frac{\gamma}{2}\coth\!\left(\frac{\omega_{0}}{2k_{B}T}\right)
	[x,[x,\rho]].
	\end{aligned}
	\label{eq:CL_rep_p}
\end{equation}%
with the effective mean-field Hamiltonian
\begin{equation}
	\begin{aligned}
	H_{\mathrm{eff}}(t)
	&= \omega_{0}a^{\dagger}a
	+ \gamma(M-1)\left\langle p(t)\right\rangle x \\
	&= \omega_{0}a^{\dagger}a
	+ 2\gamma(M-1)\operatorname{Im}\!\left\langle a(t)\right\rangle x.
	\end{aligned}
	\label{eq:Heff_CL}
\end{equation}%
where $x=a+a^{\dagger }$ and $p=(a-a^{\dagger })/i$.

The collective nature of the Caldeira--Leggett dynamics can again be traced back to the fact that all degenerate modes couple to the same environmental quadratures through the collective operators $X=A+A^{\dagger}$ and $P=(A-A^{\dagger})/i$, with $A=\sum_{\ell=1}^{M}a_{\ell}$. Expanding the dissipative and diffusive terms therefore generates crossed contributions involving different modes, in direct analogy with the collective Lindblad structure discussed in the weak-coupling regime. In particular, the common reservoir reorganizes the degenerate manifold into collective bright and dark sectors, so that dissipation acts preferentially on the bright quadrature selected by the bath. At the representative level, this structure appears through the $M$-dependent damping term governing the quadrature dynamics. Unlike the weak-coupling Markovian case, however, the collective contribution is no longer exhausted by a single enhanced exponential decay rate. Instead, the reservoir-induced damping acts directly on the phase-space quadratures, so that the collective parameter $M$ controls not only the relaxation timescale, but also the qualitative dynamical sector itself, namely whether the evolution is overdamped, critically damped, or underdamped. The emergence of multiple Mpemba-type crossings in the underdamped regime should therefore be understood as a genuinely collective consequence of the interplay between common-bath damping and coherent quadrature rotation.

This representative form clarifies what changes, and what survives,
relative to the weak-coupling Markovian description. The thermal sector is
still driven irreversibly toward the bath, but the coherent sector is no
longer exhausted by a single scalar amplitude. Instead, the damping acts
directly on the quadratures, so the collective factor $M$
controls whether the coherent relaxation is monotonic or oscillatory. This
is the structural reason why the Caldeira--Leggett regime can exhibit
delayed crossings or repeated reorderings even in the absence of reservoir
memory.

At the level of the expectation values, the representative dynamics,
following from $\left\langle \dot{x}\right\rangle =\omega _{0}\left\langle
p\right\rangle $ and $\left\langle \dot{p}\right\rangle =-\omega
_{0}\left\langle x\right\rangle -2\gamma M\,\left\langle p\right\rangle $,
retains the form of a damped harmonic oscillator 
\begin{equation}
	\left\langle \ddot{x}\right\rangle +2\gamma M\left\langle \,\dot{x}%
	\right\rangle +\omega _{0}^{2}\left\langle x\right\rangle =0,
	\label{eq:CL_damped_oscillator}
\end{equation}%
making the qualitative structure of the dynamics transparent. For $\gamma
M>\omega _{0}$, the motion is overdamped and typically monotonic. For $%
\gamma M=\omega _{0}$, it is critically damped, while for $\gamma M<\omega
_{0}$, the motion is underdamped and the quadratures exhibit damped
oscillations with frequency $\Omega =\sqrt{\omega _{0}^{2}-(\gamma M)^{2}}$.
In the underdamped sector, $\left\langle x(t)\right\rangle $ and $%
\left\langle p(t)\right\rangle $ contain oscillatory factors of the form $%
e^{-\gamma Mt}\cos (\Omega t)$ and $e^{-\gamma Mt}\sin (\Omega t)$, so the
reduced dynamics acquires a genuine rotational dynamics in the quadrature
plane, absent from the weak-coupling Markovian regime.

For the numerical illustrations below, it is useful to recast the total energy in a way that mirrors this physical picture, separating the incoherent occupation associated with the covariance matrix from the coherent occupation carried by the quadrature first moments. We denote the former by $\nu(t)$, with $\nu_0$ its initial value and $\nu_{\rm eq}$ its equilibrium value fixed by the bath temperature. The latter is denoted by $c(t)$. Thus, the total energy can be written as $E=M\omega_0[\nu(t)+c(t)]$, where $\nu(t)=\nu_{\rm eq}+(\nu_0-\nu_{\rm eq})e^{-2\gamma t}$, while $c(t) = \left\langle x(t)\right\rangle ^{2}+\left\langle p(t)\right\rangle ^{2}/{4}$ is determined by the quadrature dynamics.

The term $\nu(t)$ carries the purely thermal relaxation, while $c(t)$ keeps
track of how the initial displacement is redistributed between the two
quadratures. This decomposition provides the missing link between the
damped-oscillator picture and the Mpemba diagnostics: $\nu(t)$ measures the
incoherent drift toward the bath, whereas $c(t)$ records the coherent part
of the motion along the trajectory in the quadrature plane. Unlike the weak-coupling
Lindblad case, the coherent contribution now depends on the phase of the
initial displacement through $\left\langle x(t)\right\rangle$ and $\left\langle
p(t)\right\rangle$. In this sense, the Caldeira--Leggett regime preserves
the same competition between incoherent and coherent channels found in weak
coupling, but the coherent channel now carries an internal rotational
dynamics. This extra quadrature dynamics is precisely what allows the
Caldeira--Leggett regime to delay a crossing in the overdamped sector or
split it into multiple crossings in the underdamped sector.

This structure has several consequences that are relevant for Mpemba-type
behavior. First, the crossing time in the Caldeira--Leggett model does not,
in general, admit a simple closed expression. It is sensitive to the initial
phase of the state, may depend separately on $\Re(\alpha)$ and $\Im(\alpha)$%
, and need not be unique. Second, multiple crossings are not a generic
feature of the model as such, but arise primarily in the underdamped sector,
where quadrature oscillations imprint a micro-oscillatory structure on $c(t)$
and, more weakly, on the energy itself. By contrast, in the overdamped
sector the relaxation remains monotonic and qualitatively closer to the
weak-coupling Markovian case. In that regime the solution contains two
relaxation modes, 
\begin{equation}
	x(t)\sim A\,\mathrm{e}^{-\lambda_+ t}+B\,\mathrm{e}^{-\lambda_- t}, \qquad
	\lambda_\pm=\gamma M\pm\sqrt{(\gamma M)^2-\omega_0^2},  \label{eq:CL_lambdas}
\end{equation}
with $\lambda_-<\lambda_+$, so that the long-time dynamics is controlled by
the slower mode. In the strongly overdamped limit $\gamma M\gg\omega_0$,
this slower rate behaves as 
\begin{equation}
	\lambda_-\approx \frac{\omega_0^2}{2\gamma M}.  \label{eq:CL_lambda_slow}
\end{equation}
This slow mode is absent from the weak-coupling Markovian description and
tends to push crossing events to later times. The Caldeira--Leggett regime
therefore enriches the problem in two distinct ways: it can delay crossings
in the overdamped sector and generate multiple crossings within finite time
windows in the underdamped sector \cite%
{ChatterjeeTakadaHayakawa2024,LonghiQuantum2025}.

\subsection{Underdamped collective dynamics: representative multiple crossings}

\begin{figure}[!htb]
	\centering
	\subfloat[Energy $E$.]{%
		\includegraphics[width=0.48\linewidth]{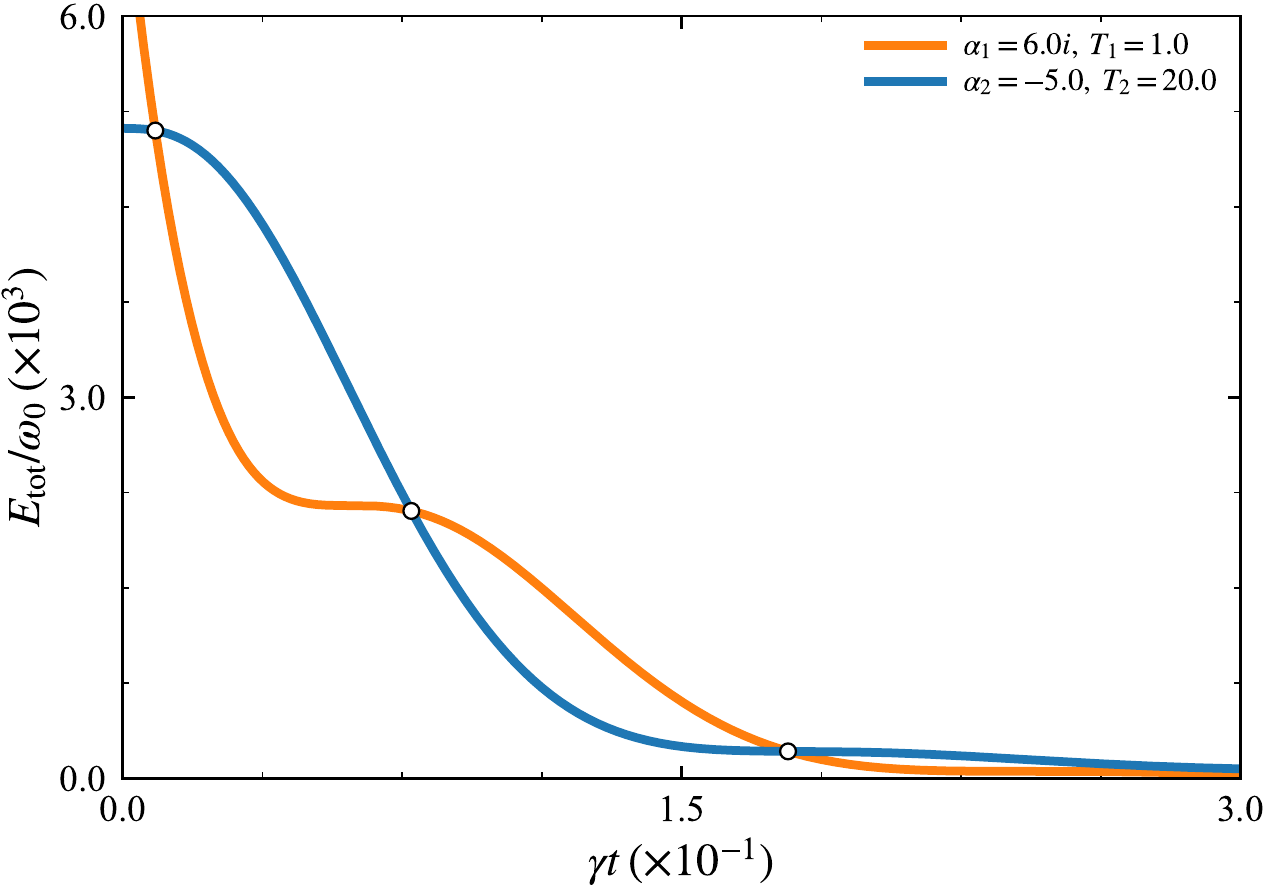}%
	}\hfill\subfloat[KL divergence.]{%
		\includegraphics[width=0.48\linewidth]{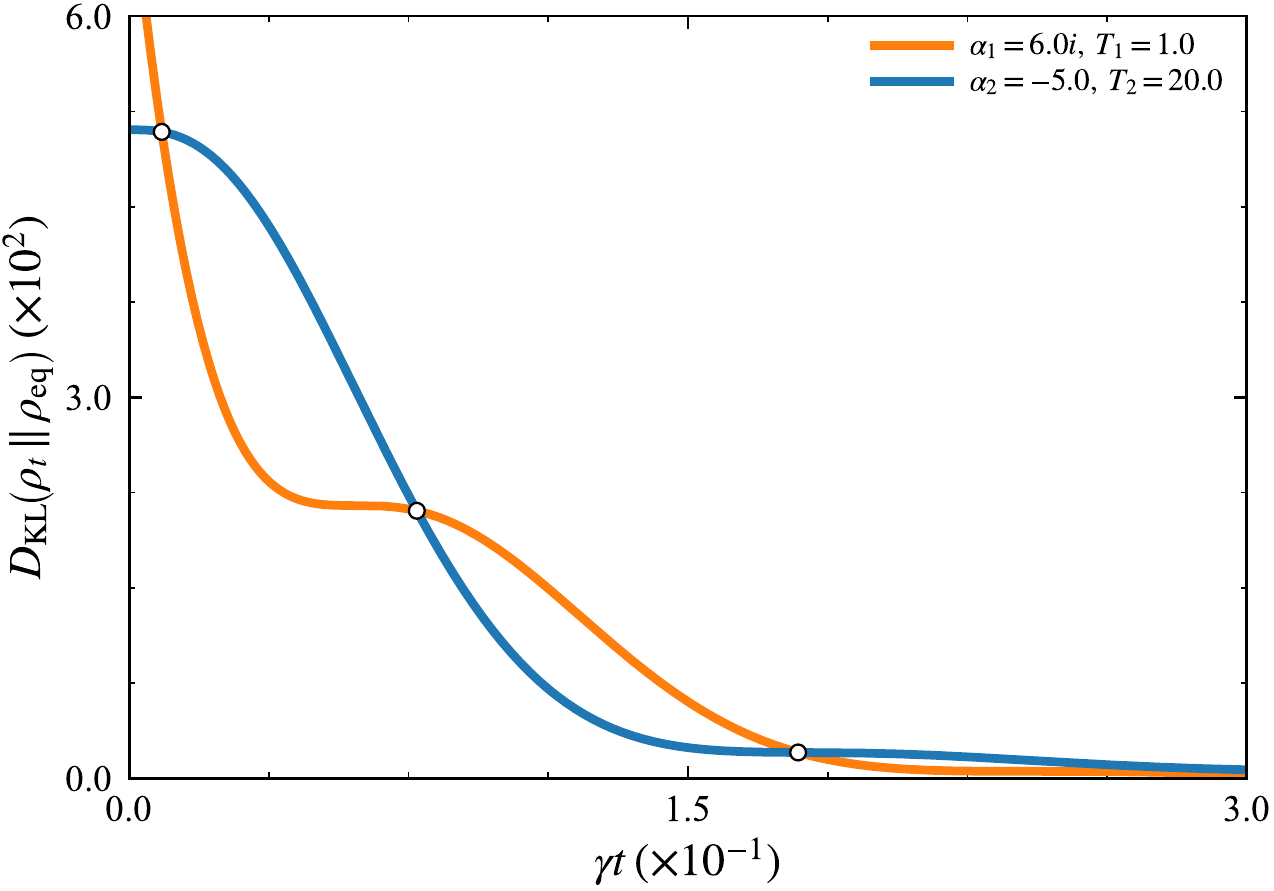}%
	}
	\caption{Underdamped collective Caldeira--Leggett dynamics for $M=10$,
		$\omega_0=20$, $\gamma=1$, $T=1.0$, $(\alpha_1,T_1)=(6i,1.0)$, and
		$(\alpha_2,T_2)=(-5.0,20.0T)$. Because $\gamma M<\omega_0$, the damped
		rotation in the quadrature plane produces three crossings in both the energy $E$
		and the KL divergence. The white markers identify the crossing times.}
	\label{fig:cl_1}
\end{figure}

The underdamped example below makes this mechanism especially transparent,
because the coherent part does not simply decay: it rotates while being
damped, so the relative ordering between different preparations can be
reshuffled more than once before the thermal contribution takes over.
Figure~\ref{fig:cl_1} shows an underdamped collective Caldeira--Leggett
example with $M=10$, $\omega_0=20$, $\gamma=1$, bath temperature $T=1.0$, and
initial preparations $(\alpha_1,T_1)=(6i,1.0)$ and $(\alpha_2,T_2)=(-5.0,20.0T)$.
Since $\gamma M/\omega_0=0.5<1$, the dynamics lies comfortably inside the
underdamped sector. The two states inject their coherent energy into
different quadratures: the first preparation is purely imaginary, whereas
the second is purely real and negative. The subsequent rotation in the quadrature plane
therefore reorganizes the relaxation hierarchy several times before the
thermal part dominates.

This structure is visible already in the energy $E$, which exhibits three
crossings over the window $0<\gamma t\lesssim 0.2$. The KL divergence shows
the same three reorderings and confirms that they are not artifacts of a
single moment of the distribution. In other words, the transient quadrature
microstructure predicted by Eq.~\eqref{eq:CL_damped_oscillator} is strong
enough to reshuffle the full number statistics. This is the cleanest message
of the strong-coupling collective regime: multiple Mpemba-type crossings can
appear even without reservoir memory, purely because damping and rotation in the
quadrature plane act together.

\FloatBarrier

\section{Degenerate bosonic modes weakly coupled to a non-Markovian reservoir}

\label{sec:nonmark}

We now turn to the genuinely non-Markovian extension, in which the
system--reservoir coupling remains weak, as in the weak-coupling Markovian
regime, but the reservoir is now allowed to have a structured spectrum and
finite memory. Within the same representative approach, the reduced dynamics
of the representative bosonic mode takes the form 
\begin{equation}
	\begin{aligned}
	\dot{\rho}(t)
	&= -i[H_{\mathrm{eff}}(t),\rho(t)] \\
	&\quad + \frac{\Gamma_-(t)}{2}
	\Big([a,\rho a^\dagger]-[a^\dagger,a\rho]\Big) \\
	&\quad + \frac{\Gamma_+(t)}{2}
	\Big([a^\dagger,\rho a]-[a,a^\dagger\rho]\Big).
	\end{aligned}
	\label{eq:NM_time_local_ME}
\end{equation}
where $\Gamma_\pm(t)$ encode reservoir memory effects. The instantaneous effective mean-field Hamiltonian retains the same nonlinear mean-field structure as in the
weak-coupling Markovian description, but with time-dependent coefficients, 
\begin{equation}
	\begin{aligned}
	H_{\mathrm{eff}}(t)
	&= \big(\omega_0+\delta(t)\big)a^\dagger a \\
	&\quad + \frac{i\Gamma_c(t)}{2}(M-1)
	\Big(\left\langle a^\dagger(t)\right\rangle a \\
	&\qquad\qquad\qquad
	- \left\langle a(t)\right\rangle a^\dagger\Big).
	\end{aligned}
	\label{eq:NM_Heff}
\end{equation}
where $\delta(t)$ is a time-dependent Lamb shift and $\Gamma_c(t)$ is the
collective coherent contribution induced by the reservoir.

Relative to Eq.~\eqref{eq:ME_GKSL}, the weak-coupling thermal structure is
preserved, and the extension can be summarized by the replacements
\begin{subequations}
	\label{eq:NM_replacements}
	\begin{align}
		\gamma(n_B+1) &\rightarrow \Gamma_-(t),
		\label{eq:NM_replacements_a}
		\\
		\gamma n_B &\rightarrow \Gamma_+(t),
		\label{eq:NM_replacements_b}
		\\
		\omega_0 &\rightarrow \omega_0+\delta(t),
		\label{eq:NM_replacements_c}
		\\
		\gamma &\rightarrow \Gamma_c(t)
		\quad \text{in the mean-field feedback term.}
		\label{eq:NM_replacements_d}
	\end{align}
\end{subequations}

In the Markovian case, both $\langle a\rangle$ and $\langle n(t)\rangle$ relax exponentially. In the non-Markovian regime, the coherent amplitude instead evolves through a generally non-exponential relaxation function determined by the bath correlations. Denoting by $g(t)$ the corresponding single-channel relaxation function, one may write $\langle a\rangle=\alpha e^{-i\phi(t)}[g(t)]^M$,
where $\phi(t)$ contains the time-dependent phase induced by the Lamb shift. Accordingly, the coherent contribution to the occupation number becomes proportional to $|g(t)|^{2M}$, while the incoherent thermal contribution relaxes through $|g(t)|^2$. The collective parameter $M$ therefore enhances the separation between coherent and incoherent relaxation channels even in the presence of reservoir memory. The mean occupation can then be decomposed into incoherent and
coherent parts, 
\begin{equation}
	\left\langle n(t)\right\rangle=n_{\mathrm{inc}}(t)+n_{\mathrm{coh}}(t),
	\label{eq:n_decomp_nonmark}
\end{equation}
where $n_{\mathrm{inc}}(t)$ describes relaxation toward the bath occupation $%
n_B$, while $n_{\mathrm{coh}}(t)$ originates from the initial displacement
amplitude.

To keep the notation aligned with the Markovian regime, we write the total
energy of the degenerate $M$-mode set simply as $E=M\omega_0\,\left\langle n(t)\right\rangle$. For two initial states $1$ and $2$ evolving in the same structured reservoir, the energy difference can be written as 
\begin{equation}
	\begin{aligned}
	\Delta E(t)
	&= M\omega_0\Big[\Delta n_0\,F_{\mathrm{inc}}(t) \\
	&\qquad\qquad
	+ \Delta\alpha\,F_{\mathrm{coh}}(t)\Big].
	\end{aligned}
	\label{eq:DeltaE_nonmark_general}
\end{equation}
where $F_{\mathrm{inc}}(t)$ and $F_{\mathrm{coh}}(t)$ are the generally
non-exponential relaxation functions associated with the incoherent and
coherent sectors. Crossings are determined by the existence of $t_{\times}$
such that $\Delta E(t_{\times})=0$, which, in the non-Markovian regime, typically must be solved numerically.
The physical mechanism, however, remains the same as in the Markovian case:
competition between incoherent thermal relaxation and coherent collective
relaxation, now dressed by memory through non-exponential relaxation
functions and a shifted relaxation timescale.

\subsubsection*{Lorentzian structured reservoir: explicit relaxation
	functions and crossing shift}

A particularly useful non-Markovian model is provided by a Lorentzian
spectral density, 
\begin{equation}
	J(\omega)=\frac{1}{2\pi}\,\frac{\gamma_0\lambda^2}{(\omega-\omega_c)^2+%
		\lambda^2},  \label{eq:Lorentzian_J}
\end{equation}
where $\gamma_0$ sets the overall system--reservoir coupling scale, $\lambda$
is the Lorentzian width and thus sets the reservoir correlation time $%
\tau_B\sim \lambda^{-1}$, and $\omega_c$ is the central frequency of the
structured reservoir. Defining also the detuning $\Delta=\omega_0-\omega_c$,
a convenient finite-temperature assumption is thermal proportionality
between decay and excitation channels, $\Gamma_{-}(t)=\kappa(t)(n_{B}+1)$ and $\Gamma_+(t)=\kappa(t)n_{B}$, so that the bath occupation remains fixed at $n_B$, while the overall
dissipative strength is governed by a single time-dependent coefficient $
\kappa(t)$ defined below. For the Lorentzian spectrum, the mode amplitude can be written as 
\begin{equation}
	\begin{aligned}
	g(t)
	&= \mathrm{e}^{-(\lambda-i\Delta)t/2}
	\left[\cosh\!\left(\frac{dt}{2}\right)\right. \\
	&\qquad\qquad\left.
	+ \frac{\lambda-i\Delta}{d}\sinh\!\left(\frac{dt}{2}\right)\right].
	\end{aligned}
	\label{eq:lorentz_g}
\end{equation}
where $d=\sqrt{(\lambda-i\Delta)^2-2\gamma_0\lambda}$ and 
$\Delta=\omega_0-\omega_c$. 
The time-dependent decay rate and Lamb shift follow from 
\begin{equation}
	\begin{aligned}
	\kappa(t)
	&= -2\,\Re\!\left(\frac{\dot g(t)}{g(t)}\right), \\
	\delta(t)
	&= -\Im\!\left(\frac{\dot g(t)}{g(t)}\right).
	\end{aligned}
	\label{eq:kappa_delta_from_g}
\end{equation}
It is convenient to define the integrated damping function 
$K(t)=\int_0^t \kappa(s)\dd s$, such that $K(t)=-\ln |g(t)|^2$. The incoherent and coherent relaxation functions then become
$F_{\mathrm{inc}}(t)=\mathrm{e}^{-K(t)}=|g(t)|^2$ and
$F_{\mathrm{coh}}(t)=\mathrm{e}^{-M K(t)}=|g(t)|^{2M}$.
Accordingly, $|\left\langle a(t)\right\rangle|^2=|\alpha|^2\,F_{\mathrm{coh}}(t)$,
$\left\langle n(t)\right\rangle=n_B+(n_0-n_B)F_{\mathrm{inc}}(t)+|\alpha|^2
F_{\mathrm{coh}}(t)$ and for two initial states $1,2$ it follows
$\Delta E(t)=M\omega_0\big[\Delta n_0\,F_{\mathrm{inc}}(t)+\Delta \alpha\,F_{\mathrm{coh}}(t)\big]$,
$\Delta n_0=n_{0,1}-n_{0,2}$, and
$\Delta \alpha=|\alpha_1|^2-|\alpha_2|^2$.
The crossing time therefore satisfies the implicit equation 
\begin{equation}
	\begin{aligned}
	&\Delta n_0\,F_{\mathrm{inc}}(t_{\times})
	+ \Delta\alpha\,F_{\mathrm{coh}}(t_{\times})=0.
	\end{aligned}
	\label{eq:crossing_lorentz}
\end{equation}
In the Markovian limit $\lambda\gg\gamma_0,|\Delta|$, one has $g(t)\to
e^{-\gamma t/2}$ and therefore $F_{\mathrm{inc}}(t)\to e^{-\gamma t}$ and $%
F_{\mathrm{coh}}(t)\to e^{-M\gamma t}$, as in the weak-coupling Markovian
case.

A useful analytic approximation for moderate memory can be obtained in the
resonant case $\Delta=0$ by approximating 
\begin{equation}
	\begin{aligned}
	\kappa(t)
	&\approx \gamma_0\big(1-\mathrm{e}^{-\lambda t}\big), \\
	K(t)
	&\approx \gamma_0\left[t-
	\frac{1-\mathrm{e}^{-\lambda t}}{\lambda}\right].
	\end{aligned}
	\label{eq:kappa_approx}
\end{equation}

The collective ingredient in the non-Markovian regime remains the same common-bath mechanism already present in the weak-coupling Markovian case. Since all degenerate modes couple to the same reservoir channel, the dynamics is still organized into bright and dark collective sectors, and the coherent contribution retains its collectively enhanced character through the $M$-dependent factor $F_{\rm coh}(t)=|g(t)|^{2M}$. Reservoir memory does not introduce an additional collective channel; rather, it dresses the same bright-sector relaxation through non-exponential damping, time-dependent rates, and the Lamb shift. As a result, non-Markovianity can delay, narrow, or split the Mpemba crossing window without changing the underlying collective origin of the effect.

Using the factorization $\Delta n_0+\Delta \alpha\,\mathrm{e}^{-(M-1)K(t_{\times})}=0$,
one obtains 
\begin{equation}
	K(t_{\times})
	= \frac{1}{M-1}\ln\!\left(-\frac{\Delta\alpha}{\Delta n_0}\right),
	\label{eq:K_crossing}
\end{equation}
with the same sign condition as in the Markovian case. Defining 
\begin{equation}
	S\equiv
	\frac{1}{(M-1)\gamma_0}
	\ln\!\left(-\frac{\Delta\alpha}{\Delta n_0}\right),
\end{equation}
Eq.~\eqref{eq:kappa_approx} yields the implicit relation $t_{\times}-(1-\mathrm{e}^{-\lambda t_{\times}})/\lambda=S$
and a simple non-Markovian (NM) first-order estimate is 
\begin{equation}
	t_{\times}^{(\mathrm{NM})}
	\approx S+\frac{1-\mathrm{e}^{-\lambda S}}{\lambda}.
	\label{eq:tx_shift}
\end{equation}
Since $(1-\mathrm{e}^{-\lambda S})/\lambda>0$, memory shifts the crossing to
later times while preserving the same underlying mechanism, namely
competition between incoherent thermal relaxation and coherent collective
relaxation.

\subsection{Memory-induced delayed crossings and narrow crossing windows}

\begin{figure}[!htb]
	\centering
	\subfloat[Energy $E$.]{%
		\includegraphics[width=0.48\linewidth]{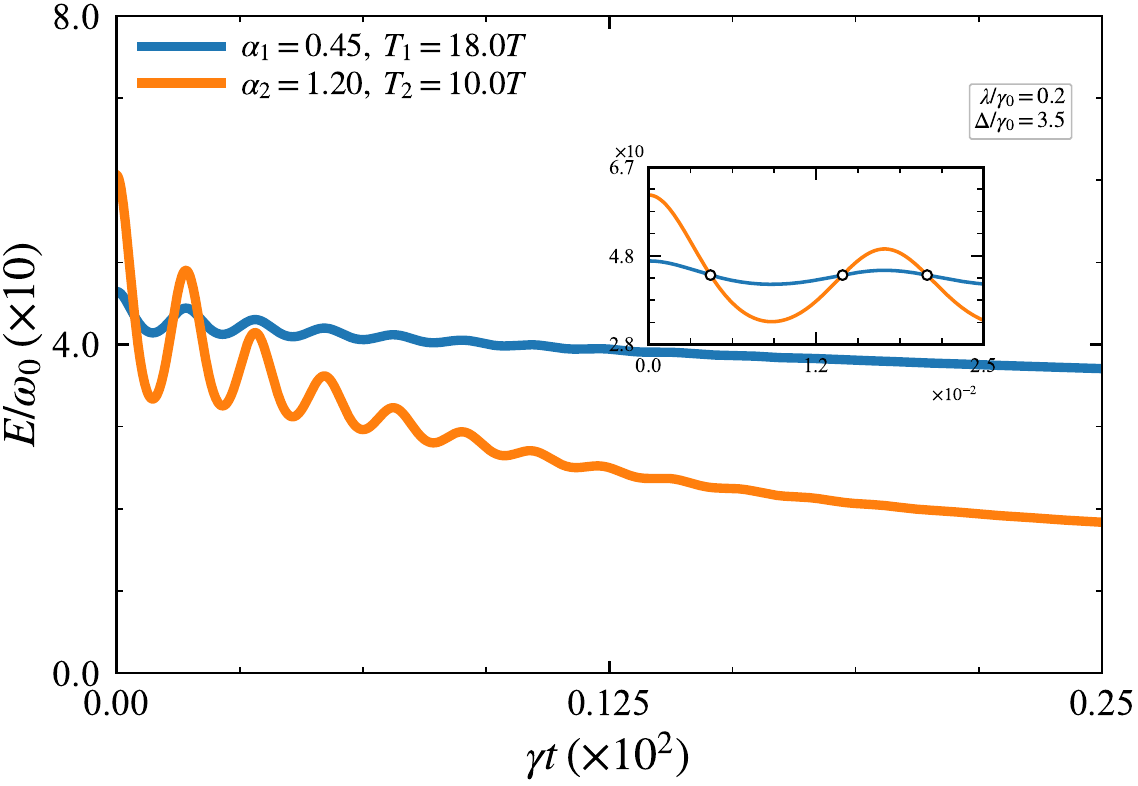}%
	}\hfill\subfloat[KL divergence.]{%
		\includegraphics[width=0.48\linewidth]{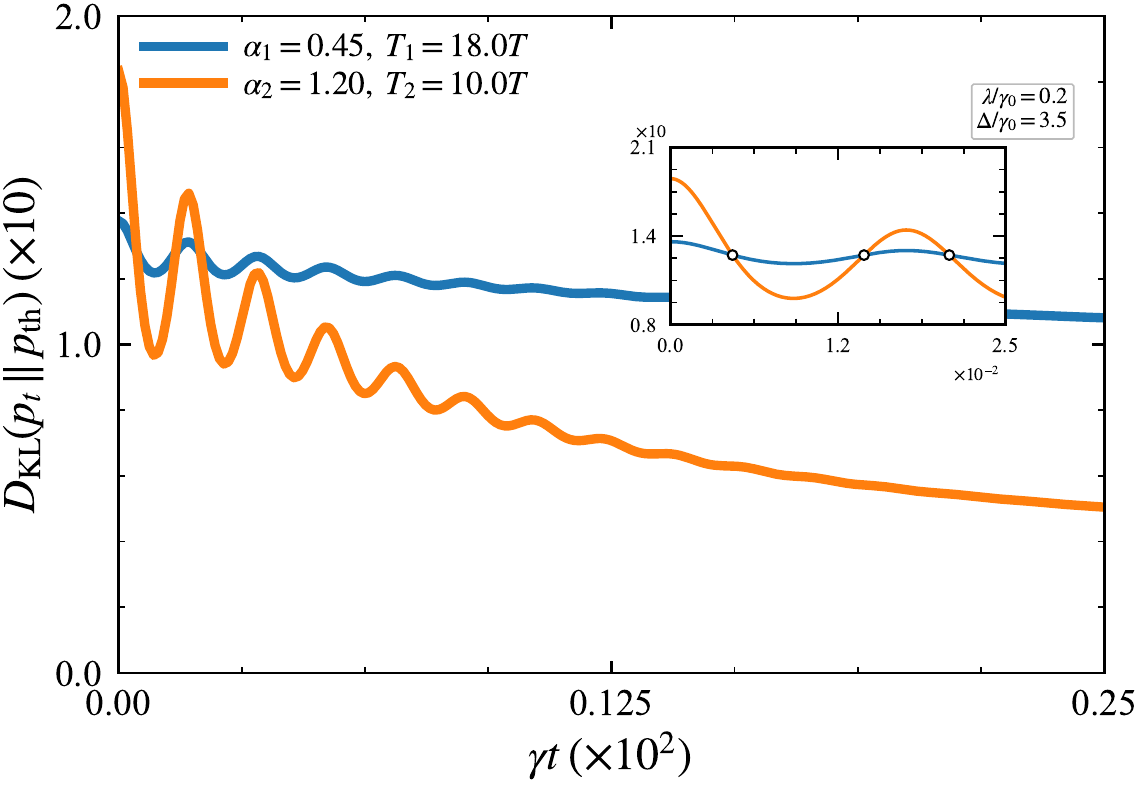}%
	}
	\caption{Lorentzian non-Markovian collective relaxation for $M=30$,
		$\omega_0=10$, $\gamma_0=1$, $T=1.0$, $\lambda/\gamma_0=0.2$,
		$\Delta/\gamma_0=3.5$, $(\alpha_1,T_1)=(0.45,18.0T)$, and
		$(\alpha_2,T_2)=(1.20,10.0T)$. The main panels show the overall delayed
		transfer of hierarchy, while the inset resolves three crossings inside a
		narrow early-time window generated by reservoir memory.}
	\label{fig:nm_1}
\end{figure}

Figure~\ref{fig:nm_1} displays a Lorentzian non-Markovian example with
$M=30$, $\omega_0=10$, $\gamma_0=1$, bath temperature $T=1.0$,
$\lambda/\gamma_0=0.2$, $\Delta/\gamma_0=3.5$, and initial preparations
$(\alpha_1,T_1)=(0.45,18.0T)$ and $(\alpha_2,T_2)=(1.20,10.0T)$. Both states
start above the bath temperature, so the basic competition is still the one
between thermal and coherent sectors. The new ingredient is that memory makes
both sectors relax through the non-exponential kernel $K(t)$, while the
finite detuning and the Lamb shift reshape the short-time dynamics.

On the scale of the main panels, the orange trajectory relaxes faster overall
and eventually becomes the closer state in both the energy and the KL
divergence. The inset reveals that this transfer of hierarchy is not a single
clean crossing. Instead, the two trajectories exchange order three times
inside a narrow early-time window. This is exactly the fine structure one
expects when the Markovian channel-competition mechanism is dressed by
memory: the crossing is postponed and split into a compact cluster of
reordering events. The KL divergence makes that narrow window especially
transparent, but the same pattern is already visible in the energy, which
shows that the delayed window is a genuine dynamical effect.

\FloatBarrier


\section{Conclusions}

We investigated how collective dissipation induced by a common thermal reservoir modifies Mpemba-type relaxation in a degenerate manifold of bosonic modes. Starting from fully symmetric multimode descriptions and employing a representative mean-field reduction, we analyzed weak-coupling Markovian, strong-coupling Markovian Caldeira--Leggett, and weak-coupling non-Markovian regimes within a unified framework. In all three cases, the essential collective ingredient originates from the common-bath coupling, which reorganizes the degenerate manifold into bright and dark sectors and introduces $M$-dependent relaxation scales absent in independent-mode dynamics.

In the weak-coupling Markovian regime, the dynamics separates naturally into incoherent thermal relaxation and a collectively enhanced coherent contribution. The latter relaxes at a rate proportional to $M\gamma$, reflecting the relaxation of the bright collective sector selected by the common reservoir. This structure makes the Mpemba mechanism analytically transparent and explains how the redistribution of the initial nonequilibrium between incoherent and coherent channels can generate Mpemba, anti-Mpemba, or criterion-dependent behavior. In particular, the collective bright-sector relaxation produces explicit $M$-dependent modifications of the crossing timescale.

In the strong-coupling Caldeira--Leggett regime, the same collective reservoir structure acquires a qualitatively richer dynamical character because dissipation acts directly on the phase-space quadratures. In this case, the collective parameter $M$ not only renormalizes relaxation rates, but also controls the dynamical sector itself, determining whether the evolution is overdamped, critically damped, or underdamped. As a consequence, overdamped trajectories tend to delay crossings, whereas underdamped trajectories can generate repeated transient reorderings and multiple Mpemba-type crossings within finite time windows. These effects arise from the interplay between collective damping and coherent quadrature rotation.

In the weak-coupling non-Markovian regime, reservoir memory preserves the same underlying collective bright-sector mechanism already present in the Markovian case, but dresses it through non-exponential relaxation, time-dependent rates, and a Lamb shift. Memory therefore does not introduce an independent collective channel; instead, it reshapes the temporal structure of the collective relaxation process, delaying, narrowing, or fragmenting the crossing window depending on the reservoir parameters.

The numerical examples based on the energy and the Kullback--Leibler divergence show that the identification of the ``faster'' preparation is not determined solely by the dynamical generator, but also by the diagnostic used to quantify proximity to equilibrium. While the energy provides an operationally transparent characterization of relaxation, the Kullback--Leibler divergence is more sensitive to the full reorganization of the occupation-number distribution and can therefore reveal transient inversions hidden at the level of the first moment alone.

Taken together, our results show that collective reservoir-induced relaxation can substantially reorganize Mpemba-type dynamics in bosonic systems. Depending on the dynamical regime, the common bath may accelerate coherent relaxation through bright-sector enhancement, generate quadrature-induced transient reorderings, or reshape the crossing structure through memory effects. More broadly, the present framework highlights how collective dissipation and reservoir structure jointly control nonequilibrium relaxation hierarchies in open quantum systems.

\section*{Acknowledgments}
The authors would like to thank CAPES and FAPESP for support.

\bibliographystyle{apsrev4-2}
\bibliography{manuscript}

\end{document}